\documentclass[aps,prd,twocolumn,preprintnumbers,nofootinbib,superscriptaddress,amsmath]{revtex4-1}
\usepackage{graphicx}
\usepackage{amsmath}
\usepackage{amssymb}
\usepackage{amsfonts}
\usepackage{hyperref}
\usepackage{url}
\usepackage{color}
\usepackage{bm}
\usepackage{array}

\newcommand{\be}{\begin{equation}}
\newcommand{\ee}{\end{equation}}
\newcommand{\ba}{\begin{eqnarray}}
\newcommand{\ea}{\end{eqnarray}}
\newcommand{\nn}{\nonumber}

\def\lcdm{$\Lambda$CDM }

\makeatother

\begin{document}

\preprint{IFT-UAM/CSIC-21-78}

\title{Testing the $\Lambda$CDM paradigm with growth rate data and machine learning}

\author{Rub\'{e}n Arjona}
\email{ruben.arjona@uam.es}
\affiliation{Instituto de F\'isica Te\'orica UAM-CSIC, Universidad Auton\'oma de Madrid,
Cantoblanco, 28049 Madrid, Spain}

\author{Alessandro Melchiorri}
\email{alessandro.melchiorri@roma1.infn.it}
\affiliation{Physics Department and INFN, Università di Roma “La Sapienza”, Ple Aldo Moro 2, 00185, Rome, Italy}

\author{Savvas Nesseris}
\email{savvas.nesseris@csic.es}
\affiliation{Instituto de F\'isica Te\'orica UAM-CSIC, Universidad Auton\'oma de Madrid,
Cantoblanco, 28049 Madrid, Spain}

\date{\today}

\begin{abstract}
The cosmological constant $\Lambda$ and cold dark matter (CDM) model ($\Lambda\text{CDM}$) is one of the pillars of modern cosmology and is widely used as the de facto theoretical model by current and forthcoming surveys. As the nature of dark energy is very elusive, in order to avoid the problem of model bias, here we present a novel null test at the perturbation level that uses the growth of matter perturbation data in order to assess the concordance model. We analyze how accurate this null test can be reconstructed by using data from forthcoming surveys creating mock catalogs based on $\Lambda\text{CDM}$ and three models that display a different evolution of the matter perturbations, namely a dark energy model with constant equation of state $w$ ($w$CDM), the Hu \& Sawicki and designer $f(R)$ models, and we reconstruct them with a machine learning technique known as the Genetic Algorithms. We show that with future LSST-like mock data our consistency test will be able to rule out these viable cosmological models at more than 5$\sigma$, help to check for tensions in the data and alleviate the existing tension of the amplitude of matter fluctuations $S_8=\sigma_8\left(\Omega_m/0.3\right)^{0.5}$.
\end{abstract}
\maketitle

\section{Introduction}
Cosmology has entered into a precision era due to the abundance of high precision observational data acquired over the last decades leading to the construction of the standard \lcdm model \cite{Aghanim:2018eyx}, with a cosmological constant ($\Lambda$) and a cold dark matter (CDM) component. Given the simplicity and low number of free parameters (just six in the minimal $\Lambda$CDM) Bayesian analyses have shown that the \lcdm is preferred over a plethora of other alternative models \cite{Heavens:2017hkr}. 

Even though the spatially flat \lcdm model is widely used as the de facto theoretical model by current and forthcoming surveys, it still remains a phenomenological model, since we ignore what is the nature of dark matter (DM) and dark energy (DE) and with the disadvantage that there exists a growing discordance in some cosmological parameters using different observations \cite{Perivolaropoulos:2021jda}, thus suggesting that the \lcdm scenario might be an approximation to a more fundamental theory that remains currently unreachable \cite{DiValentino:2020vhf}. 

In fact, it is worth mentioning the discrepancy between the Hubble constant $H_0$ obtained through the distance ladder and the one derived through analyses of the CMB, see Ref.~\cite{DiValentino:2021izs} for a recent review. Also, recent analyses of the Planck 2018 data suggest the possibility of a small, but non-zero, curvature, thus implying a non-flat universe, see Refs.~\cite{Aghanim:2018eyx,DiValentino:2019qzk,Handley:2019tkm}. This deviation from flatness could be due to unaccounted for systematic errors or due to a statistical fluctuation \cite{DiValentino:2020srs} and clearly deserves further investigation. Let us also point out a reported $\sim 4\sigma$ deviation from the \lcdm model of the dark energy equation of state $w(z)$ making use of quasars at redshifts up to $z\sim 7.5$  \cite{Risaliti:2018reu}. 

Moreover, the \lcdm prediction of the amplitude of matter fluctuations defined as $S_8 \equiv \sigma_8\left(\Omega_m/0.3\right)^{0.5}$, where $\sigma_8$ is the root mean square of matter fluctuations on a $8h^{-1}\text{Mpc}$ scale, coming from the Planck CMB data (under $\Lambda$CDM) is about $2-3 \sigma$ higher than the direct estimation coming from cosmic shear measurements (see, e.g.
\cite{DiValentino:2020vvd}).

Although these tensions or discrepancies mentioned could be related to unaccounted for systematic errors, there also exists the attractive alternative of new physics in the form of modified gravity (MG) or DE models. Given the plethora of MG and DE models in the literature, substantial endeavors have recently been placed to furnish a unified framework which encloses some of these models like the Effective Field Theory (EFT) \cite{Gubitosi:2012hu,Hu:2013twa} or the Effective Fluid Approach (EFA) \cite{Arjona:2018jhh,Arjona:2019rfn,Arjona:2020gtm,Cardona:2020ama}. In fact, due to the wide range of alternative models it is difficult for the observations to interpret the results on the cosmological parameters since they depend on the particular model assumed.

Since the nature of DE is very elusive and not well understood, to circumvent this problem there is growing interest in non-parametric reconstruction methods and model-independent approaches \cite{Nesseris:2010ep} that overcome the biases of having to define a certain theoretical model. In this regard, machine learning (ML) algorithms have provided innovative solutions for extracting information in a theory agnostic manner \cite{Ntampaka:2019udw}. Some of these algorithms have been applied to reconstructions of model-independent tests, i.e. using a function that only depends on observed quantities and not on any theoretical model. These null tests are useful to check for possible tensions and systematics in the data, or to probe for hints of new physics. 

A main advantage of the null tests is that any deviations at any redshift from the expected value imply the failure of the  assumptions made \cite{Marra:2017pst}. Null tests have been already applied to the concordance \lcdm model \cite{Sahni:2008xx,Zunckel:2008ti,Nesseris:2010ep}, interacting DE models \cite{vonMarttens:2018bvz}, the growth-rate data \cite{Nesseris:2014mfa,Marra:2017pst,Benisty:2020kdt}, the cosmic curvature \cite{Yahya:2013xma,Cai:2015pia,Benisty:2020otr,Li:2014yza}, to probe the scale-independence of the growth of structure in the linear regime \cite{Franco:2019wbj} and the homogeneity of the Universe \cite{Arjona:2021hmg}. 

On the other hand, MG theories can be properly modified such that they can mimic the evolution of the expansion of different dark energy models like the \lcdm but behaving differently at the perturbation level, like for example the $f(R)$ designer models \cite{Arjona:2018jhh} or the Hordenski designer (HDES) \cite{Arjona:2019rfn}. These models would be indistinguishable when using geometric probes, while at the same time dark energy might evolve with time, leading to dark energy clustering or have a non-adiabatic component \cite{Arjona:2020yum}. Hence there exists a degeneracy between these models at the background level, but that could be in principle broken using dynamic probes like the growth rate data which traces the matter density perturbations.

The main purpose of the present analysis is to present a new consistency test for the \lcdm model at the perturbations level which could be evaluated from the growth rate data. Assuming a homogeneous and isotropic Universe, from the equation of the growth of matter density contrast which is described as a second order differential equation, we construct a null test with the potential of verifying the aforementioned assumptions used to derive the evolution of the matter density contrast by using direct observations and without needing to specify a model. Our novel null test has the advantage that it does not contain higher derivative terms, which makes the error increase when noisy data are used, hence delivering tighter constraints for the \lcdm model. It is also quite generic and has to be valid at all redshifts. We show that with a survey like the LSST, the growth rate data will be able to discriminate a wide range of MG theories from $\Lambda$CDM. 

The reconstructions of the LSST-like mock data are performed using a particular ML algorithm known as the Genetic Algorithms (GA). This is a stochastic minimization and symbolic regression algorithm and has the advantage of avoiding the issue of biases since it is a non parametric method that allows us to make the least number of assumptions concerning the underlying cosmology.

\begin{figure}[!t]
\centering
\hspace*{-4mm}
\includegraphics[width = 0.57\textwidth]{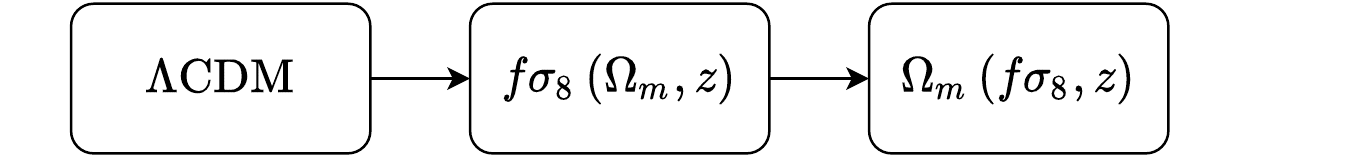}
\caption{A flowchart describing the creation of the null test presented in Section.~\ref{sec:null}. The end goal is to express the matter density $\Omega_m$ as a function of $f\sigma_8(z)$ and the redshift $z$ via the Lagrange inversion theorem.  \label{fig:flow}}
\end{figure}

The structure of the paper is as follows. In Sec.~\ref{sec:null} we properly define our \lcdm consistency test for the matter density dubbed $\textrm{Om}_{f\sigma_8}(z)$ and in Sec.~\ref{sec:models} we describe the theoretical models used in the analysis. Following, in Sec.~\ref{sec:data} we describe the data and the LSST-like mocks used and in Sec.~\ref{sec:GA} explain the Machine Learning reconstruction algorithm used in the analysis, known as the Genetic Algorithms. Finally, in Sec.~\ref{sec:results} we present our results and in Sec.~\ref{sec:conclusions} we summarize our conclusions.
\section{The null test\label{sec:null}}
As depicted in the flowchart of Fig.~\ref{fig:flow}, our end goal is to express the matter density parameter $\Omega_m$ as a function of $f\sigma_8(z)$ and the redshift $z$, i.e $\textrm{Om}(f\sigma_8,a)$. Note that the redshift $z$ and the scale factor $a$ are related as $a=\frac{1}{1+z}$. A cosmological probe that is not geometric in origin is the growth function of the linear matter density contrast defined as $\delta_m \equiv \frac{\delta \rho_m}{\rho_m}$. The advantage of these measurements comes from the fact that the growth of matter density perturbations is mostly induced by the motion of matter and then it is very sensitive to both any modified gravity that deviates from GR and the expansion of the Universe $H(a)$ \cite{Nesseris:2013fca}. Assuming a homogeneous and isotropic
universe and neglecting neutrinos, then the growth factor
$\delta_{m}(a)$ in MG theories satisfies the following differential equation on subhorizon scales $k^2 \gg a^2H^2$
{\small
\begin{equation}\label{eq:growth}
\delta_{m}^{\prime \prime}(a)+\left(\frac{3}{a}+\frac{H^{\prime}(a)}{H(a)}\right) \delta_{m}^{\prime}(a)-\frac{3}{2} \frac{\Omega_{\mathrm{m}, 0}G_\mathrm{eff}(a)/G_\mathrm{N}}{a^{5} H(a)^{2} / H_{0}^{2}} \delta_{m}(a)=0,
\end{equation}} 
where the primes indicate differentiation with respect to the scale factor $a$. It is clear that when $G_\mathrm{eff}(a)/G_\mathrm{N}=1$ we recover GR, while in general MG models $G_\mathrm{eff}$ can be dependent on time and scale \cite{Amendola:2007rr,Tsujikawa:2007gd,Nesseris:2008mq,Nesseris:2009jf}.
The growth of matter perturbation $\delta_m(a)$ in GR assuming flatness,  considering a constant dark energy equation of state $w$ and neglecting radiation can be expressed as \cite{Buenobelloso:2011sja}
\begin{equation}
\delta(a)=a\;{}_{2} F_{1}\left(-\frac{1}{3 w}, \frac{1}{2}-\frac{1}{2 w} ; 1-\frac{5}{6 w} ; a^{-3 w}\left(1-\Omega_{m}^{-1}\right)\right),
\end{equation}
where ${}_{2} F_{1}(a, b ; c ; z)$ is a hypergeometric function, see Ref.~\cite{Abramowitz:1974} for more details. Assuming the $\Lambda\text{CDM}$ model, where $w=-1$, we can define the quantity $\Delta(a)$ as
\ba
\Delta(a)&=&\frac{\delta_m(a)}{\delta_m(1)}\nn\\
&=&\frac{a~{}_2F_1\left[\frac{1}{3},1,\frac{11}{6},a^3\left(1-\Omega^{-1}_{m}\right)\right]}{{}_2F_1\left[\frac{1}{3},1,\frac{11}{6},\left(1-\Omega^{-1}_{m}\right)\right]}.
\label{eq:sigma}
\ea
To derive the actual $\textrm{Om}(f\sigma_8,a)$ test we first do a series expansion on Eq.~(\ref{eq:sigma}) around $\Omega^{-1}_{m}=1$ and keep the first $15$ terms. Then, we apply the Lagrange inversion theorem\footnote{The Lagrange inversion theorem asserts that given an analytic function, one can estimate the Taylor series expansion of the inverse function. In other words, given the function $y=f(x)$, where $f$ is analytic at a point $p$ and $f'(p)\neq 0$, the theorem allows one to solve the equation for $x$ and write it as a power series $x=g(y)$, see \cite{Abramowitz:1974}.} to invert the series expansion and to write the inverse matter density $\Omega^{-1}_{m}$ as a function of $\Delta(a)$, i.e $\Omega^{-1}_{m}\equiv \textrm{Om}^{-1}_{f{\sigma_8}}\equiv \textrm{Om}^{-1}_{f{\sigma_8}}(a,\Delta)$. We used the inverse matter density as we found it was more numerically stable and robust. For example, the first two terms of the  expansion are
\begin{equation}
   \textrm{Om}^{-1}_{f{\sigma_8}}(a,\Delta)=1+\frac{11\left(\Delta-a\right)}{2a\left(1-a^3\right)}+\cdots,
\end{equation}
then the actual $\textrm{Om}(f\sigma_8,a)$ null test is given by
\begin{equation}\label{eq:nulltest}
    \textrm{Om}(f\sigma_8,a)=\frac{1}{\textrm{Om}^{-1}_{f_{\sigma_8}}(a,\Delta)}.
\end{equation}
Note that we have considered the series expansion of the combination $\frac{\delta_m(a)}{\delta_m(1)}$ (see Eq.~(\ref{eq:sigma})), instead of only $\delta_m(a)$, and again the reason behind it is because that approach was found to be more numerically stable and robust.

It is important to note that what is measurable is not exactly the growth $\delta_{m}(a)$, but the combination $f\sigma_8(a)\equiv f(a)\cdot \sigma_8 (a)$, where $f(a)\equiv \frac{d\ln \delta_m(a)}{d \ln a}$ is the growth rate of structure and $\sigma_8(a)=\sigma_8\frac{\delta_m(a)}{\delta_m(1)}$ is the redshift-dependent rms fluctuations of the linear
density field. The value of $f\sigma_8(a)$ can be obtained from the ratio of the monopole to the quadrupole of the redshift-space power spectrum, which depends on the parameter $\beta=f/b_0$, where $b_0$ is the bias. The advantage of $f\sigma_8(a)$ is that it is independent of the bias \cite{Song:2008qt}.
Performing direct manipulations of the definition of $f\sigma_8(a)$ one can show that (see Ref.~\cite{Arjona:2020kco} for more details) the quantity $\Delta(a)$ used in our $\textrm{Om}(f\sigma_8,a)$ null test can be written in terms of $f\sigma_8(a)$ as
\begin{equation}
    \Delta(a)\equiv \frac{\delta_m(a)}{\delta_m(1)}=\frac{1}{\sigma_8}\int_0^a \frac{f\sigma_8(x)}{x} dx, \label{eq:fs8}
\end{equation}
and through the definition of $f\sigma_8$ we can also derive the useful relation
\be\label{eq:sigma8}
\sigma_{8} =\int_{0}^{1} \frac{f \sigma_{8}(x)}{x} d x,
\ee
hence the function $\Delta(a)$ of Eq.~(\ref{eq:fs8}) can be derived solely having a reconstructed function for $f\sigma_8(a)$. In essence, if we reconstruct the function $f\sigma_8(a)$ we can have a reconstructed function for $\Delta(a)$ using Eq.~(\ref{eq:fs8}) and use it to have a model independent consistency test of the matter density for the \lcdm model. In other words, at all redshifts we should have that $\textrm{Om}(f\sigma_8,a)=\Omega_{m}$ and any deviation from $\Omega_{m}$ could be due to various reasons
\begin{itemize}
    \item Tensions in the growth rate data.
    \item Detection of modified gravity and $G_\mathrm{eff}(a)/G_\mathrm{N}\neq 1$.
    \item A presence of shear or strong dark energy perturbations.
    \item Deviations from the FLRW metric.
\end{itemize}

Similarly, we could use our reconstruction of $f\sigma_8(a)$ to create a null test for the amplitude of matter fluctuations $S_8$, which is another parameter that also quantifies the matter fluctuations, and is defined as
\be
S_8=\sigma_8\sqrt{\Omega_{m}/0.3},
\ee
since we can obtain $\sigma_8$ through Eq.~(\ref{eq:sigma8}) and $\Omega_m$ is derived from the null test defined in Eq.~(\ref{eq:nulltest}). Hence we define  this other null test as $\textrm{O}_{S_{8}}(f\sigma_8,a)$, which is defined as
\ba
\textrm{O}_{S_{8}}(f\sigma_8,a)&\equiv&\sigma_8\sqrt{\textrm{Om}(f\sigma_8,a)/0.3}\nn\\
&=&\left(\int_{0}^{1} \frac{f \sigma_{8}(x)}{x} d x\right) \sqrt{\textrm{Om}(f\sigma_8,a)/0.3}.~~\label{eq:Os8}
\ea
The new null test of Eq.~\eqref{eq:Os8} is a function only of the $f\sigma_8(a)$ GA reconstruction and is constant only for the \lcdm model. 

\section{The models\label{sec:models}}
We will create LSST-like simulated data based on four different cosmological models defined below to test our $\textrm{Om}(f\sigma_8,z)$ and $\textrm{O}_{S_{8}}(f\sigma_8,z)$ null test to first, estimate how much the errors on the null test will be with a similar LSST survey and second, to inspect the validity and generality of these null tests.

\subsection{The \lcdm model}
We will assume a \lcdm model, i.e. $w=-1$, with $\Omega_{m}=0.3$ and $\sigma_8=0.8$.
\subsection{The $w$CDM model}
For the $w$CDM model we will consider two cases, one with $w=-1.09$, which is in the range of $\sim 3\sigma$ with the best-fit value of Planck 2018, and another with $w=-1.2$, which is somewhat more extreme, so as to examine how well our tests work. In both cases we will further assume  $\Omega_{m}=0.3$ and $\sigma_8=0.8$ as the \lcdm model. In this scenario the Hubble equation is given by
\be
H(a)^{2} / H_{0}^{2}=\Omega_{\mathrm{m}} a^{-3}+\left(1-\Omega_{\mathrm{m}}\right) a^{-3(1+w)},
\ee
which for $w=-1$ reduces to the expression for the \lcdm model. 

Interestingly, if $w\neq -1$, then dark energy is able to cluster and the scale when this effect can occur depends on the properties of the fluid such as the pressure perturbation $\delta p$ which is related to the sound speed and the anisotropic stress $\sigma$, see Ref.~\cite{Sapone:2009mb} for more details. In Eq.~(\ref{eq:growth}) we have to take into account DE perturbations if DE can cluster at sufficiently small scales as explained in Ref.~\cite{Nesseris:2014qca}. To consider this effect in our mock catalog we modify Eq.~(\ref{eq:growth}) by implementing the function
\be\label{eq:Qeff}
Q(a)=1+\frac{1-\Omega_{m}}{\Omega_{m}} \frac{1+w}{1-3 w} a^{-3 w},
\ee
which is going to act as a modified Newton's constant, i.e. $G_\mathrm{eff}(a)/G_\mathrm{N}\equiv Q(a)$. In this case, Eq.~(\ref{eq:Qeff}) is derived assuming there is no anisotropic stress and zero DE sound speed as shown in Ref.~\cite{Sapone:2009mb}.

\subsection{The $f(R)$ models}
Following Ref.~\cite{Arjona:2018jhh} the  action for $f(R)$ models can be expressed as
\be
S=\int d^{4} x \sqrt{-g}\left[\frac{1}{2 \kappa} f(R)+\mathcal{L}_{m}\right],
 \ee
where $\mathcal{L}_{m}$ is the Lagrangian of matter, $\kappa=8\pi G_\mathrm{N}$ is a constant and $G_\mathrm{N}$ is Newton's constant. Varying the action with respect to the metric, we arrive at the well known field equations 
\be
F G_{\mu \nu}-\frac{1}{2}(f(R)-R F) g_{\mu \nu}+\left(g_{\mu \nu} \square-\nabla_{\mu} \nabla_{\nu}\right) F=\kappa T_{\mu \nu}^{(m)},
\ee
where $F=f'(R)$. In the sub-horizon approximation, i.e., when the modes are deep in the horizon $(k^2\gg a^2H^2)$, the Newtonian potentials,  using the equations of motion, can be written as
 \ba
\Psi &=&-4 \pi G_\mathrm{N} \frac{a^{2}}{k^{2}} \frac{G_\mathrm{eff}}{G_\mathrm{N}} \bar{\rho}_{m} \delta_{m}, \\
\Phi &=&-4 \pi G_\mathrm{N} \frac{a^{2}}{k^{2}} Q_\mathrm{eff} \bar{\rho}_{m} \delta_{m},
 \ea
where the functions $G_\mathrm{eff}/G_\mathrm{N}$ and $Q_\mathrm{eff}$ that can depend on time and scale are described as
\ba\label{eq:geff}
G_\mathrm{eff} / G_\mathrm{N} &=&\frac{1}{F} \frac{1+4 \frac{k^{2}}{a^{2}} \frac{F_{R}}{F}}{1+3 \frac{k^{2}}{a^{2}} \frac{F_{R}}{F}}, \\
Q_\mathrm{eff} &=&\frac{1}{F} \frac{1+2 \frac{k^{2}}{a^{2}} \frac{F_{R}}{F}}{1+3 \frac{k^{2}}{a^{2}} \frac{F_{, R}}{F}}.
\ea
 
\subsubsection{The Hu $\&$ Sawicki model}

The well known Hu $\&$ Sawicki (HS) model \cite{Hu:2007nk} is described by the lagrangian
\begin{equation}
\label{Hu}
f(R)=R-m^2 \frac{c_1 (R/m^2)^n}{1+c_2 (R/m^2)^n},
\end{equation}
which may be rewritten, after some algebraic manipulations as \cite{Basilakos:2013nfa}
\be
\label{Hu1}
f(R)= R- \frac{2\Lambda }{1+\left(\frac{b \Lambda }{R}\right)^n},
\ee
where $\Lambda= \frac{m^2 c_1}{2c_2}$ and $b=\frac{2 c_2^{1-1/n}}{c_1}$. In Ref.~\cite{Basilakos:2013nfa} the authors found that when written in the form given by Eq.~\eqref{Hu1}, it is clear to infer why the HS model satisfies all the solar system tests. In essence,  if $b \to 0$ \lcdm is recovered, and  if $b \to \infty$ a matter dominated universe is obtained  i.e.,
\ba
 \lim_{b\rightarrow0}f(R)&=&R-2\Lambda , \nn \\
\lim_{b\rightarrow \infty}f(R)&=&R.
\ea

If the parameter $b$ is sufficiently small, the HS model can be considered as a ``perturbation'' around the \lcdm model. This is important as we do not assume the usual approximation of fixing the background to \lcdm when analyzing the HS model. However, solving numerically the equation for the Hubble parameter is not trivial so we follow a different approach. Instead of approximating the background, we solve the field equations and use an approximate, accurate analytical expression for the Hubble parameter. Following Ref.~\cite{Basilakos:2013nfa} it can be shown that the Hubble parameter for the HS model can be written as
\ba
\label{eq:HSHubbleApproximation}
H_\mathrm{HS}(a)^2 = H_{\Lambda}(a)^2 + b_\mathrm{hs} \, \delta H_1(a)^2 + b_\mathrm{hs}^2\, \delta H_2(a)^2 + \cdots,~~~~~~
\ea
which is an analytical approximation that works extremely well, e.g. for $b\leq 0.1$ the average error with respect to the numerical solution is $10^{-5}\%$ for redshifts $z\leq 30$. 

For our mock catalog we modify Eq.~(\ref{eq:growth}) by implementing the corresponding $G_\mathrm{eff}$ function as defined in Eq.~(\ref{eq:geff}) for the HS model and with the proper Hubble rate, see Eq.~(\ref{eq:HSHubbleApproximation}). We will also assume the following parameters $\Omega_{m}=0.3, k=300 H_{0}, b=10^{-4}$ and $\sigma_{8,0}=0.8$. As can be seen from Fig.~\ref{fig:omfs8} with the parameters mentioned above, our null test deviates strongly from \lcdm when using the HS model as the fiducial cosmology. However, recent observations can constrain the $b$ parameter significantly, see for example Ref.~\cite{Cardona:2020ama} where the authors found that $b<10^{-8}$, but when using this value our null test behaves exactly as \lcdm and both models then would not be distinguishable.

\subsubsection{The $f(R)$ Designer model}
There is a particular class of $f(R)$ models that behave as the \lcdm model at the background, while manifesting differences in the evolution of the linear perturbations.
These models are known as the designer $f(R)$ models \cite{Multamaki:2005zs,delaCruzDombriz:2006fj,Nesseris:2013fca}, which we will dubbed as DES-fR from now on. The DES-fR model satisfying all viability conditions (see  for instance Ref.~\cite{Pogosian:2007sw}) is given by \cite{Nesseris:2013fca}
\ba
\label{des}
f(R)&=&R-2\Lambda+\alpha~H_0^2\left(\frac{\Lambda }{R-3 \Lambda }\right)^{c_{0}} \times \nn \\
& & {}_2F_1\left(c_{0},\frac{3}{2}+c_{0},\frac{13}{6}+2c_{0},\frac{\Lambda }{R-3 \Lambda }\right)\;,
\ea
where $c_{0}=\frac{1}{12} \left(-7+\sqrt{73}\right)$, $\alpha$ is a free dimensionless parameter, $H_0$ is the Hubble constant, $\Lambda$ is a constant, and ${ }_{2} F_{1}(a, b ; c ; z)$  is a hypergeometric function.

For our mock catalog we modify Eq.~(\ref{eq:growth}) by implementing the corresponding $G_\mathrm{eff}$ function as defined in Eq.~(\ref{eq:geff}) for the DES-fR model. We will also assume the following parameters $\Omega_{m}=0.3, k=300 H_{0}, b=10^{-4}$ and $\sigma_{8,0}=0.8$, see Fig.~\ref{fig:omfs8}. The 
parameter $b$ was also constrained to $b<10^{-8}$ on Ref.~\cite{Cardona:2020ama}. When using the latter value we find that this model can also be detected with our null test and data coming from a survey like LSST, see Fig.~\ref{fig:omdes}.

\section{Mock Data\label{sec:data}}
The Legacy Survey of Space and Time (LSST), performed by the Vera C. Rubin Observatory \cite{Abell:2009aa}, can complement other future growth surveys and extend the probed redshift range. In our analysis we create mock LSST-like growth rate data for $f\sigma_8(z)$ based on the models mentioned above in Sec.~\ref{sec:models} to test our $\textrm{Om}(f\sigma_8,z)$ and $\textrm{O}_{S_{8}}(z)$ null test. This will allow us to study the validity and the generality of our consistency test and see how the errors will be with a future LSST-like survey. 

Being more interested in checking the $\textrm{Om}(f\sigma_8,z)$ and $\textrm{O}_{S_{8}}(z)$ test rather than concerned about systematics in the data we evaluate the growth uniformly distributed in the range $z \in[0,2]$ divided into $10$ equally spaced binds of step $dz=0.2$. The $f\sigma_8(z_i)$ function was estimated as its theoretical value from the different cosmological models plus a gaussian error (which can be either negative or positive) and assigning an error of $1\%$ of its value, which is in agreement with a similar setup to LSST accuracy as described in Refs.~\cite{Huterer:2013xky,Abell:2009aa}.

\section{Genetic Algorithms\label{sec:GA}}
The Genetic Algorithms (GA) fall under a class of machine learning methods useful for non-parametric reconstruction of data. They are build on the concept of grammatical evolution, as indicated by the genetic operations of crossover and mutation. In specific, the GA imitate the principle of evolution by the implementation of the principle of natural selection, where a group of individuals evolves as time passes by under the pressure of the stochastic operators of mutation, particularly a random change in an individual, and crossover, i.e.
the merger of different individuals to form offspring. 

The probability that a member of the population will produce offspring, or in other words, its success in reproducing, is
assumed to be proportional to its fitness. In essence, the fitness measures how accurately each individual of the population fits the data, and it is quantified through a $\chi^2$ statistic, which in our case as we are reconstructing LSST-like $f\sigma_8(z)$ data it will be given by 
\be
\chi^2=\sum^{N}_{i=1}\left(\frac{f\sigma_{8,i}-f\sigma_\mathrm{8,GA}(z_i)}{\sigma_i}\right)^2.
\ee
Our reconstruction of the growth rate
data with the GA is as follows. First, an initial population of
functions is randomly selected in order that every member of the population has an initial guess for $f\sigma_8(z)$. At this stage we also impose the following physical priors. We assume that the Universe at early times went through a phase of matter domination $(z\sim 1000)$, which implies that the linear growth behaves as $\delta_m(a) \simeq a$ at high redshifts, however we don't assume a DE model. 
Then, each member’s fitness is computed through a $\chi^2$
statistic, using as input the mock $f\sigma_8$ data. Later, the operators of mutation and crossover are
applied to the best-fitting functions in each generation, chosen
via the tournament selection, see Ref.~\cite{Bogdanos:2009ib} for
more details. This process is repeated thousands of times in order to assure convergence and with different random seeds to don't to bias the results due to a specific choice of the random
seed.
After the GA has converged, the final output is a continuous and differentiable function of redshift that
describe the $f\sigma_8$. Finally,
to provide an estimate of the errors on the reconstruct function we make use of
an analytical approach developed by Refs.\cite{Nesseris:2012tt,Nesseris:2013bia}, where the errors are obtained through a path integral over the whole functional space that can be scanned by the
GA at $1\sigma$. This GA path integral approach has been extensively tested by
\cite{Nesseris:2012tt,Nesseris:2013bia} and found to be in excellent
agreement with other error estimates methods like bootstrap Monte-Carlo.

To summarize, the GA can reconstruct any cosmological function, for example the $f\sigma_8(z)$ that we consider here, by applying
the algorithm to any dataset of choice. The advantage is that no assumptions on the particular
cosmological model or the behaviour of DE need to be made,
hence the results are model independent. For the numerical implementation of the GA used in this paper we have used the publicly available code made by one of the authors \footnote{\href{https://github.com/RubenArjona}{https://github.com/RubenArjona}}. We want to stress that besides performing a large number of GA
runs with different random seed numbers, we have also demanded
that all reconstructed functions, as well as their derivatives, are
continuous in the range of redshifts we consider, thus avoiding
spurious reconstructions and overfitting.

It is worth mentioning that the GA have been applied in the field of cosmology for several reconstructions on a wide range of data, see for example Refs.~\cite{Bogdanos:2009ib,Nesseris:2010ep, Nesseris:2012tt,Nesseris:2013bia,Sapone:2014nna,Arjona:2020doi,Arjona:2020kco,Arjona:2019fwb,Arjona:2021hmg,Arjona:2020skf,Arjona:2020axn,Arjona:2021zac,Aizpuru:2021vhd}. Other applications of the GA have been used for particle physics \cite{Abel:2018ekz,Allanach:2004my,Akrami:2009hp}, astronomy and astrophysics \cite{wahde2001determination,Rajpaul:2012wu,Ho:2019zap}. Finally, other symbolic regression methods implemented in physics and cosmology can be found at \cite{Udrescu:2019mnk,Setyawati:2019xzw,vaddireddy2019feature,Liao:2019qoc,Belgacem:2019zzu,Li:2019kdj,Bernardini:2019bmd,Gomez-Valent:2019lny}.

\section{Results\label{sec:results}}

\begin{figure*}[!thp]
\centering
\includegraphics[width = 0.48\textwidth]{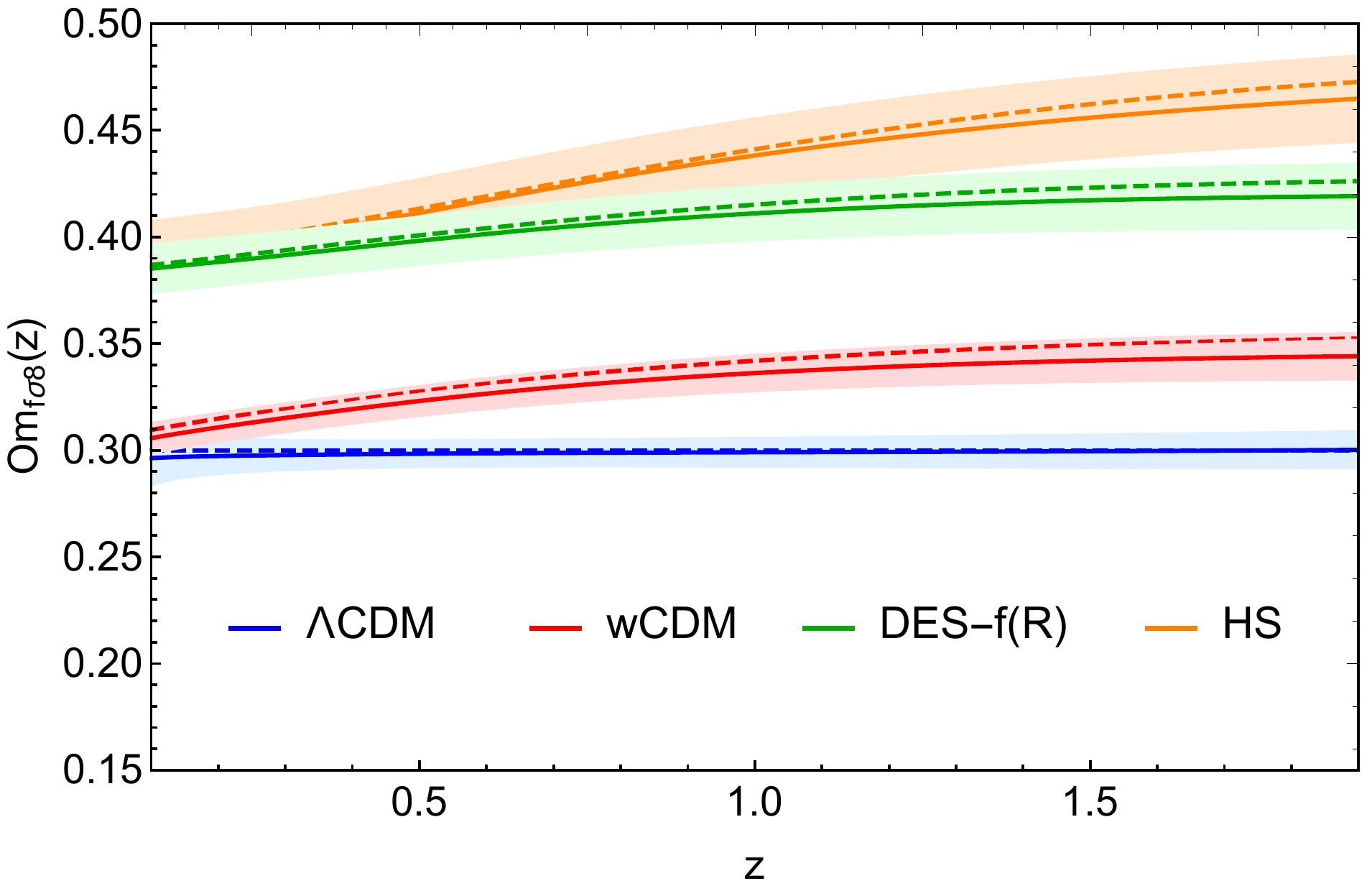}
\includegraphics[width = 0.48\textwidth]{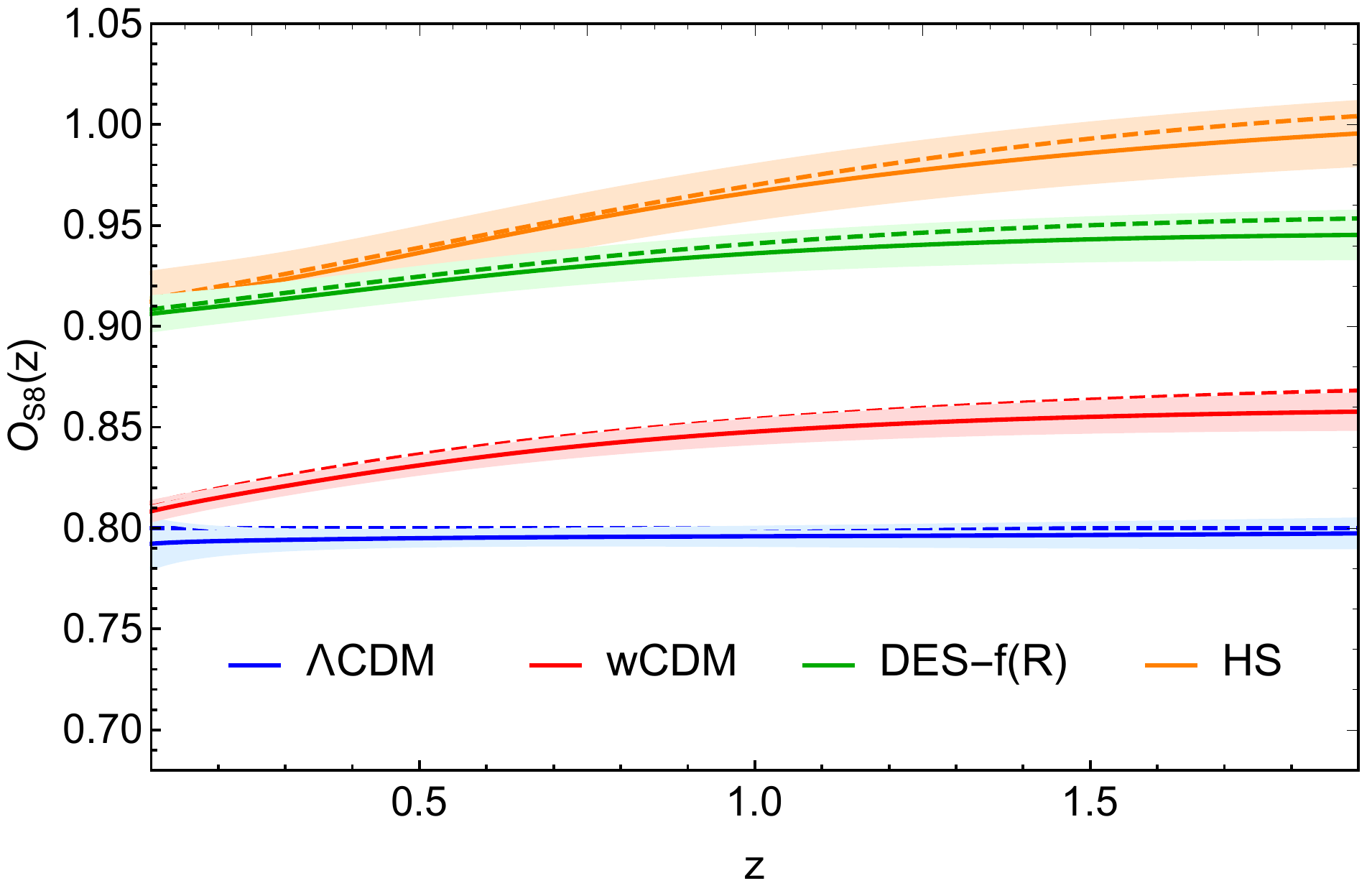}
\caption{The results for the $\mathrm{Om}_{f\sigma_8}(z)$ null test (left panel) and the $\mathrm{O}_{S_{8}}(z)$ test (right panel) along with the 1$\sigma$ errors (shaded regions) using LSST-like mock data. In both figures the fiducial \lcdm model, $w$CDM, designer $f(R)$ and Hu \& Sawicki model are represented by the blue, red, green and orange dashed-lines respectively. Also our GA reconstructions for the $\Lambda$CDM, $w$CDM, designer $f(R)$ and Hu \& Sawicki model are given by the blue, red, green and orange solid line respectively. In all cases we see that the GA recover well the best-fit value of all the models. \label{fig:omfs8}}
\end{figure*}

\begin{figure*}[!thp]
\centering
\includegraphics[width = 0.48\textwidth]{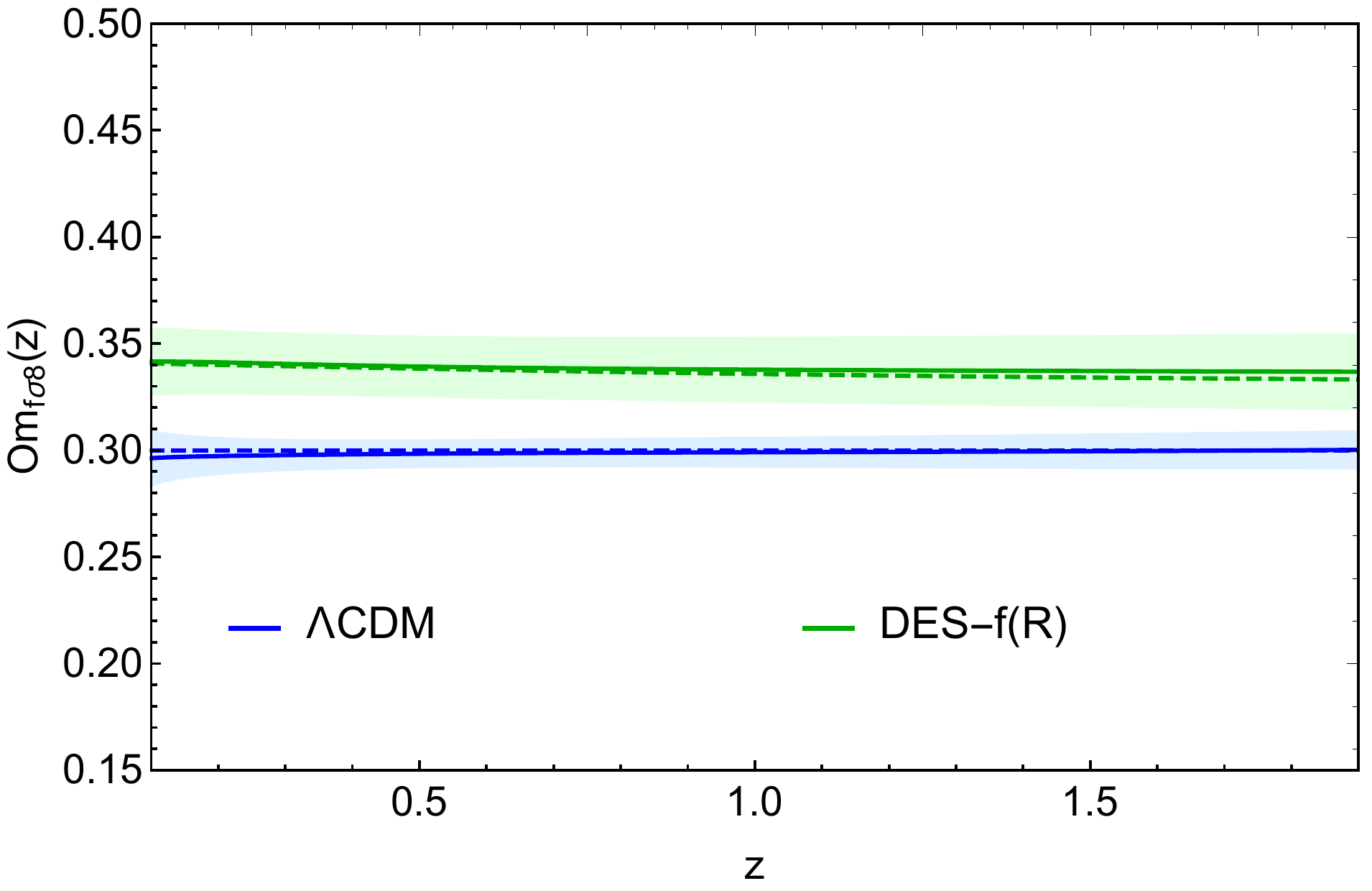}
\includegraphics[width = 0.48\textwidth]{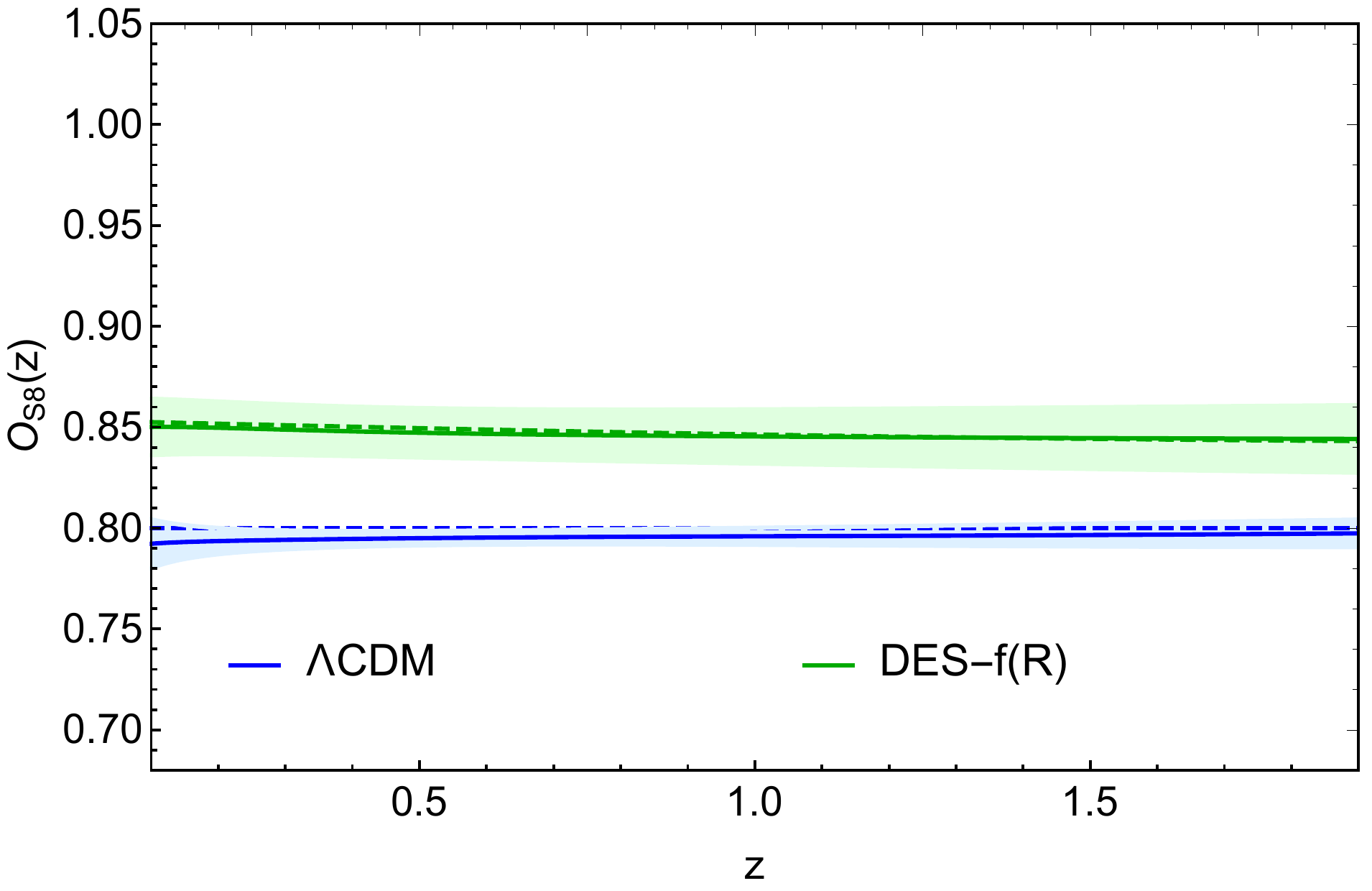}
\caption{The results for the $\mathrm{Om}_{f\sigma_8}(z)$ null test (left panel) and the $\mathrm{O}_{S_{8}}(z)$ test (right panel) along with the 1$\sigma$ errors (shaded regions) using LSST-like mock data. In both figures the fiducial \lcdm model is represented by the blue dashed-line and the designer $f(R)$ with the green dashed-line respectively. Our GA reconstructions for the \lcdm and designer $f(R)$ are given by the blue and green solid line respectively. In all cases we see that the GA recover well the best-fit value of all the models. For the designer $f(R)$ model we have used in our mocks the tight constraint of $b<10^{-8}$ and still we see a deviation of $\sim 2\sigma$ from the \lcdm model. \label{fig:omdes}}
\end{figure*}

\begin{figure*}[!thp]
\centering
\includegraphics[width = 0.48\textwidth]{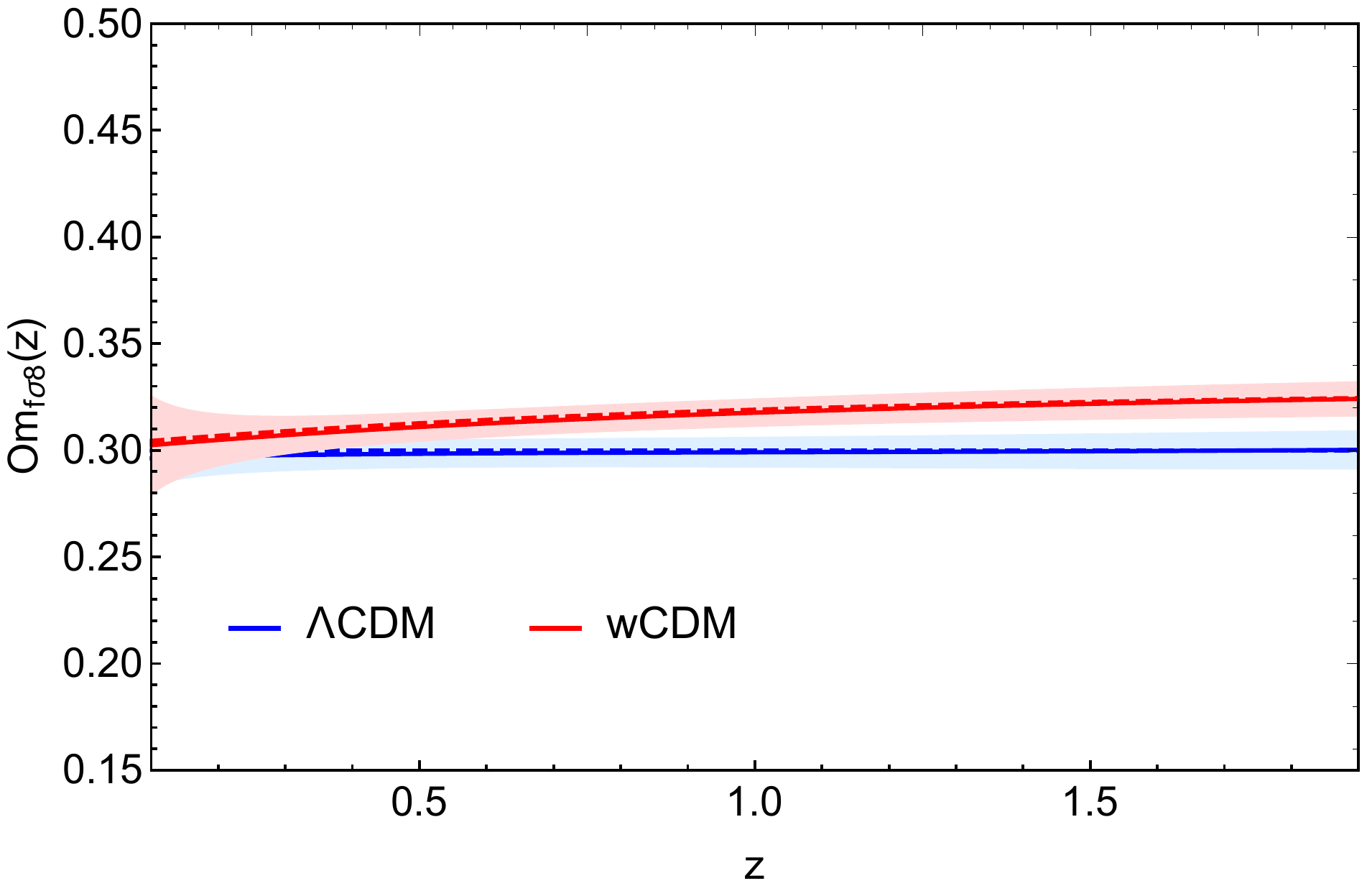}
\includegraphics[width = 0.48\textwidth]{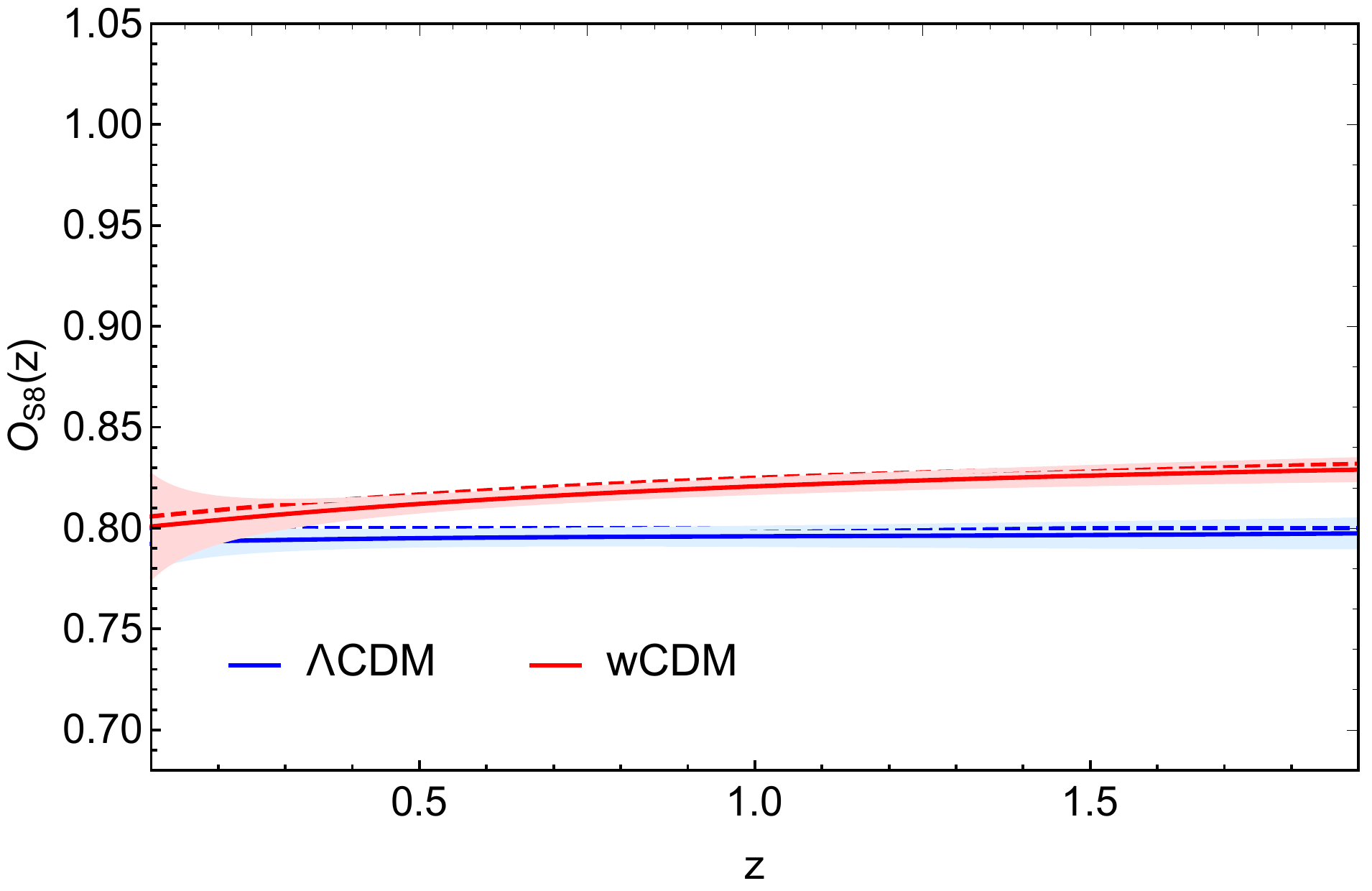}
\caption{The results for the $\mathrm{Om}_{f\sigma_8}(z)$ null test (left panel) and the $\mathrm{O}_{S_{8}}(z)$ test (right panel) along with the 1$\sigma$ errors (shaded regions) using LSST-like mock data. In both figures the fiducial \lcdm model is represented by the blue dashed-line and the $w$CDM model with the red dashed-line respectively. Our GA reconstructions for the \lcdm and $w$CDM model are given by the blue and red solid line respectively. In all cases we see that the GA recover well the best-fit value of all the models. For the $w$CDM model we have used in our mocks the $\sim 3\sigma$ constrain of $w=-1.09$ and still we see a deviation of $\sim 2-3\sigma$ from the \lcdm model for either of the tests. \label{fig:omwcdm}}
\end{figure*}

In this section we will evaluate how well the null tests are going to be reconstructed with future LSST-like data. Our goal is to perform a direct test of the standard cosmological model with the least of assumptions. We emphasize that with our null tests, any deviation from $\textrm{Om}(f\sigma_8,z)=\Omega_m=0.3$  and $\textrm{O}_{S_{8}}(z)=0.8$ which are the default values selected in our mocks, it would imply a breakdown of either the principal assumptions of the \lcdm model or that the dark energy models do not describe well the data. In Fig.~\ref{fig:omfs8} we display the results for the $\mathrm{Om}_{f\sigma_8}(z)$ null test (left panel) and the $\mathrm{O}_{S_{8}}(z)$ test (right panel) along with the 1$\sigma$ errors (shaded regions) using LSST-like mock data. 

In both figures the best-fit \lcdm model, $w$CDM, designer $f(R)$ and Hu \& Sawicki model are represented by the blue, red, green and orange dashed-lines respectively. Also our GA reconstructions for the $\Lambda$CDM, $w$CDM, designer $f(R)$ and Hu \& Sawicki model are given by the blue, red, green and orange solid line respectively. In all cases we see that the GA recover well the best-fit value of all the models. In essence, our consistency test is able to rule out these viable cosmological models at more than 5$\sigma$ for the HS and the DES $f(R)$ model and at $\sim 3\sigma$ for the $w$CDM model. 

For more realistic values of the $b$ parameter of the HS and the DES $f(R)$ models, i.e. $b<10^{-8}$, we find that the HS model is indistinguishable from $\Lambda$CDM. However, as shown in Fig.~\ref{fig:omdes}, for the DES $f(R)$ and using $b=10^{-8}$ in our mock sample we see a deviation of $\sim 2\sigma$ from the \lcdm model. In Fig.~\ref{fig:omdes} we find the results for the $Om_{f\sigma_8}(z)$ null test (left panel) and the $O_{S_{8}}(z)$ test (right panel) along with the 1$\sigma$ errors (shaded regions) using LSST-like mock data. In both figures the best-fit \lcdm model is represented by the blue dashed-line and the designer $f(R)$ with the green dashed-line respectively. Our GA reconstructions for the \lcdm and designer $f(R)$ are given by the blue and green solid line respectively. In all cases we see that the GA recover well the best-fit value of all the models. 

Finally, in Fig.~\ref{fig:omwcdm} we show the results for the $\mathrm{Om}_{f\sigma_8}(z)$ null test (left panel) and the $\mathrm{O}_{S_{8}}(z)$ test (right panel) along with the 1$\sigma$ errors (shaded regions) using LSST-like mock data for a more realistic value of the equation of state for the $w$CDM model, namely $w=-1.09$ which is $\sim 3\sigma$ away from the best-fit value derived by Planck 2018. In both figures the fiducial \lcdm model is represented by the blue dashed-line and the $w$CDM model with the red dashed-line respectively. Our GA reconstructions for the \lcdm and $w$CDM model are given by the blue and red solid line respectively. In all cases we see that the GA recover well the best-fit value of all the models. We still see a deviation of $\sim 2\sigma$ from the \lcdm model.

\section{Conclusions\label{sec:conclusions}}

In summary, we have presented a new consistency test at the perturbation level that uses the growth of matter perturbation data. In particular, assuming the $\Lambda$CDM model we apply the Lagrange inversion theorem in the solution of the equation of the growth of matter density contrast which is described as a second order differential equation over in order to obtain a conserved quantity that can be written in terms of the measurable quantity $f\sigma_8(z)$.

In order to forecast how well our new test, given by Eq.~\eqref{eq:nulltest}, can constrain deviations from the $\Lambda$CDM model, we created mock datasets based on specifications of the LSST survey and using the \lcdm model for the fiducial cosmology, and three models that display a different evolution of the matter perturbations, namely  the $w$CDM model, the Designer $f(R)$ and Hu \& Sawicki model. This approach allows us to quantify any deviations using realistic scenarios.

Then, to reconstruct the $\textrm{Om}(f\sigma_8,z)$ null test given by Eq.~\eqref{eq:nulltest} from the mock data, we use the machine learning approach, namely the GA, as this will allow us to obtain non-parametric and theory agnostic reconstructions of the data, in the form of $f\sigma_8(z)$, that we can in turn use to reconstruct $\textrm{Om}(f\sigma_8,z)$. Following this approach, we find that the GA with the  $\textrm{Om}(f\sigma_8,z)$ statistic can correctly predict the underlying fiducial cosmology at all redshifts covered by the data, as seen in Fig.~\ref{fig:omfs8} and can easily rule out several realistic modified gravity models at more than 5$\sigma$.

To conclude, we explicitly show that with a future survey like LSST, which is not only going to provide us with growth rate data with a higher quality, but also with more data points, thus helping to provide stringent constraints on modified gravity theories. It will have the required improvement to discriminate the aforementioned models in our analysis from \lcdm. Overall, the novelty of the results presented show that by very minimal assumptions on the nature of dark energy future surveys would have the capability to confirm or falsify the \lcdm model at the perturbation level.\\

\section*{Acknowledgements}
R.A. and S.N. acknowledge support from the Research Project PGC2018-094773-B-C32 and the Centro de Excelencia Severo Ochoa Program SEV-2016-0597. S.~N. also acknowledges support from the Ram\'{o}n y Cajal program through Grant No. RYC-2014-15843. A.M. acknowledges support from the Iniziativa Specifica INFN, TASP.\\

Numerical Analysis Files: The Genetic Algorithm code presented in this analysis can be found at \href{https://github.com/RubenArjona}{https://github.com/RubenArjona}.\\

\begin{appendix} 

\end{appendix}

\bibliography{fs8}

\end{document}